\documentstyle[12pt]{article}
\begin{document}

\tolerance=5000

\def\be{\begin{equation}}
\def\ee{\end{equation}}
\def\beq{\begin{eqnarray}}
\def\eeq{\end{eqnarray}}
\def\nn{\nonumber \\}
\def\e{{\rm e}}

\def\SEH{S_{\rm EH}}
\def\SGH{S_{\rm GH}}
\def\AdS5{{{\rm AdS}_5}}
\def\dS5{{{\rm dS}_5}}
\def\S4{{{\rm S}_4}}
\def\gfv{{g_{(5)}}}
\def\gfr{{g_{(4)}}}
\def\SC{{S_{\rm C}}}
\def\RH{{R_{\rm H}}}

%%%%%%%%%%%%%%%%%%%%%%%%%%%%%%%%%%%%%%%%
\thispagestyle{empty}

\makeatletter
\renewcommand{\theequation}{\thesection.\arabic{equation}}
\@addtoreset{equation}{section}
\makeatother

\

\vskip -2.1cm

\  \hfill
\begin{minipage}{3.5cm}
YITP-02-59 \\
September 2002 \\
hep-th/0209242 \\
\end{minipage}

\begin{center}

{\large\bf Casimir effect in de Sitter
and Anti-de Sitter braneworlds}

\

{\sc Emilio Elizalde$^\heartsuit$}\footnote{On leave from: IEEC/CSIC,
Edifici Nexus, Gran Capit\`a 2-4, 08034 Barcelona, Spain; \\ \hspace*{6mm}
 email:
elizalde@math.mit.edu, \ elizalde@ieec.fcr.es},
{\sc Shin'ichi Nojiri$^\spadesuit$}\footnote{email:
snojiri@yukawa.kyoto-u.ac.jp,
nojiri@nda.ac.jp}, \\
{\sc Sergei D. Odintsov$^\clubsuit$}\footnote{email:
odintsov@mail.tomsknet.ru} and
{\sc Sachiko Ogushi$^\diamondsuit$}\footnote{JSPS fellow,
email: ogushi@yukawa.kyoto-u.ac.jp}

\

$\heartsuit$ Department of Mathematics,  Massachusetts Institute
of Technology\\ 77 Massachusetts Avenue, Cambridge, MA 02139-4307

$\spadesuit$ Department of Applied Physics,
National Defence Academy \\
Hashirimizu Yokosuka 239, JAPAN

$\clubsuit$ Lab. for Fundamental Study \\
Tomsk Pedagogical University, 634041 Tomsk, RUSSIA \\

$\diamondsuit$ Yukawa Institute for Theoretical Physics,
Kyoto University, Kyoto 606-8502, JAPAN \\

\vspace*{0.5cm}

{\bf ABSTRACT}

\end{center}
We discuss the bulk Casimir effect (effective potential) for a conformal
or massive scalar when the bulk represents five-dimensional AdS or dS
space with two or one four-dimensional dS brane, which may correspond to
our universe.  Using zeta-regularization, the interesting conclusion
is reached, that for both bulks in the one-brane limit the effective
potential corresponding to the massive or to the conformal scalar
is zero. The radion potential in the presence of quantum corrections
is found. It is demonstrated that both the dS and the AdS braneworlds
may be stabilized by using the  Casimir force only.
A brief study indicates that bulk quantum effects are relevant for brane
cosmology, because they do deform the de Sitter brane.
They may also provide a natural mechanism yielding a decrease of the
four-dimensional cosmological constant on the physical brane of
the two-brane configuration.

\


\newpage

\setcounter{page}{1}
\tableofcontents

\newpage

\section{ Introduction \label{Intr}}

If our world is really multi-dimensional, as M-(string) theory predicts,
then one of the most economical possibilities for its realization is the
braneworld paradigm. Indeed, in the case when string theory
is taken in its exact vacuum state, with the five-dimensional
(asymptotically) Anti-de Sitter (AdS) sector, in a full
ten-dimensional space, the corresponding effective
five-dimensional theory represents some (gauged)
supergravity. Adding the four-dimensional surface terms
predicted by the AdS/CFT correspondence to such five-dimensional AdS
(super)gravity, one arrives at the dynamical four-dimensional
boundary (brane) of this five-dimensional manifold.
Depending on the structure of the surface terms, the choice of (bulk
and brane) matter, the assumptions about the general structure of the
brane and bulk manifold, fields content, etc., our
four-dimensional universe can be realized in a particular way as such
a brane. Brane universe can be consistent with observational data
even when the radius of the extra dimension is quite significant.
Moreover, the braneworld point of view of our universe may bring
about a number of interesting mechanisms to resolve such  well-known
problems as the cosmological constant and the hierarchy problems.

As the braneworld corresponds to a five-dimensional (usually AdS)
manifold with a four-dimensional dynamical boundary, it is clear
that, when the five-dimensional QFT is considered, the non-trivial vacuum
energy (Casimir effect, see e.g. \cite{milton} for a recent review)
should appear. Moreover, when the brane QFT is considered,
the non-trivial brane vacuum energy also appears.
The bulk Casimir effect should conceivably play a quite remarkable role in
the construction of the consistent braneworlds.
Indeed, it gives contribution to both the brane and the bulk
cosmological constants.
Hence, it is expected that it may help in the resolution
of the cosmological constant problem.

For consistency, the five-dimensional braneworld should be
stabilized (radion stabilization) \cite{GLW}, and the challenging idea
is that a very fundamental quantity, the bulk vacuum
energy (Casimir contribution), may be used explicitly for
realizing the radion stabilization. This has been checked
 in a number of models
\cite{GPT1}--%SSS,BMNO,GNS,HPP,NOO3,IZR,FMT,TK,NOS,NS,AAS,
\cite{GKV1}, although
mainly with flat branes only.  An interesting connection between
the bulk Casimir effect and supersymmetry breaking in
braneworld \cite{GP} or moving branes \cite{MT} also exists.
On the other hand, the brane Casimir effect may be used for
a braneworld realization \cite{HHR} of the anomaly-driven
(also called Starobinsky) inflation \cite{starobinsky}.

The works mentioned above discuss mainly the Casimir effect in the situation
when the brane is flat space. But also the situation in which the
brane is more realistic, say a de Sitter (dS) universe, has been
discussed in Refs.~\cite{SSS,NS}. It has been shown there
that, in an AdS bulk, the Casimir energy for the bulk conformal scalar
field in a one-brane configuration is zero.
However, in situations where the bulk is different, a non zero
contribution of the Casimir energy  is not excluded and
even a possibility may exist of gravity trapping on the brane itself.

In the present work we study the bulk Casimir effect for a conformal
or massive scalar when the bulk is a five-dimensional AdS or a dS space
and the brane is a four-dimensional dS space.
We show that zeta-regularization techniques at its full power  \cite{eez1}
can be used in order to calculate the
bulk effective potential in such braneworlds, in a quite general setting.
One interesting result we got is that, for both bulks (AdS and dS) under
discussion with one brane,
the bulk effective potential is zero for a conformal as well as for a
massive
scalar.
Applications of our results to the stabilization of the radion and to the
brane dynamics are presented as well.

The paper is organized as follows. The next section is devoted to the
discussion of a general effective potential (Casimir effect) for bulk
conformal scalar on AdS when the brane is a de Sitter space. The small
distance behavior is investigated and the one-brane limit of the potential,
 which turns out to be zero, is worked out.
As an application, we discuss the role of the leading term of the effective
potential to the brane dynamics. It is shown here that the Casimir force
only slightly deforms the shape of the 4-dimensional sphere S$_4$. The
radion
potential (in two limits), with account of the Casimir term, is found
and the stabilization of the braneworld is discussed. Using an explicit
short
distance expansion for the effective potential,
it is demonstrated that the brane may
indeed be stabilized  using the Casimir force only.

In Sect. 3 similar questions are investigated for a conformal scalar
when the brane is S$_4$,  and bulk is a five-dimensional dS space.
It is interesting that the effective potential turns out to be the same as
in the case of the previous section (AdS). Also, the one-brane limit of
effective potential is again zero. From the study of brane dynamics it turns
out that the role of the Casimir force is again that of
inducing  some deformation of the S$_4$ brane (especially close to the
poles).

In Sect. 4 the effective potential for a  massive scalar (also
with scalar-gravitational coupling) is presented, for both a dS and an
AdS bulk, when the brane is S$_4$. The small and large mass limits are
found.
The one-brane limit of the potential is again zero, even in the massive
case,
but the main non-zero correction to this limit is obtained explicitly.
Brane  stabilization due to the Casimir force for a massive scalar is
discussed
when the bulk is five-dimensional dS.

In Sect. 5 the potential for a  massive scalar without a
scalar-gravitational coupling is briefly studied for
 dS and AdS braneworlds. It is shown that it is again zero in the
one-brane limit. Finally, a short summary and an outlook are presented
in Sect. 6.

\section{The Casimir effect for a de Sitter brane \\
in a five-dimensional Anti-de Sitter background \label{Sec1}}

\subsection{Effective potential for the brane \label{Sec1.1}}

In this section, we review the calculation
of the effective potential for a de Sitter (dS) brane in a five-dimensional
anti-de Sitter (AdS) background, following Refs.~\cite{GPT1,SSS,NS}.
First, we start with the action for a conformally invariant
massless scalar with scalar-gravitational coupling,
\beq
\label{act}
{\cal S} = {1\over 2}\int d^{5}x \sqrt{g}
\left[ -g^{\mu\nu}\partial_{\mu} \phi \partial_{\nu} \phi +
\xi_{5}R^{(5)} \phi^2 \right] \; ,
\eeq
where $\xi_{5}=-3/16$, $R^{(5)}$ being the five-dimensional scalar
curvature.
This action is conformally invariant under
the conformal transformations:
\footnote{Note that there is a relation between $\alpha $ and $\beta$,
namely $- {D-2 \over 4}\alpha =\beta$, and $\xi_{D}$
depends on the dimensions as $-{ D-2 \over 4(D-1)}$,
for the general $D$-dimensional bulk.}
\beq
\label{tr1}
g_{\mu\nu} = e^{\alpha \sigma(x^{\mu})} \hat{g_{\mu\nu} } \, \quad
\phi =  e^{\beta \sigma (x^{\mu})} \hat{\phi } \; ,
\eeq
where $-{3\over 4} \alpha = \beta$.

Let us recall the expression for the Euclidean metric of the
five-dimensional AdS bulk:
\beq
\label{met1}
ds^2 &=& g_{\mu\nu}dx^{\mu}dx^{\nu}
={l^2 \over \sinh^2 z} \left( dz^2 + d\Omega _{4}^2 \right),\\
\label{ss4}
d\Omega_{4}^2 &=& d\xi^2 + \sin^2 \xi d\Omega_3^2 \; ,
\eeq
where $l$ is the AdS radius which is related to the cosmological constant of
the AdS bulk, and $d\Omega_3$ is the metric on the 3-sphere.
Two dS branes, which are  four-dimensional spheres,
are placed in the AdS background.
If we put one brane at $z_1$, which is fixed, and
the other brane at $z_2$, the distance between the branes is given by
 $L=|z_{1}- z_{2}|$.  When  $z_{2}$ tends
to $\infty$, namely $L = \infty$, the two-brane
configuration becomes a one brane configuration.

We can see that the action, Eq.~(\ref{act}), is conformally invariant under
the conformal transformations for the metric Eq.~(\ref{met1}) and
the scalar field, which are given by
\beq
\label{conf}
g_{\mu\nu} = \sinh^{-2} z\; l^{2} \hat{g}_{\mu\nu}\; , \quad
\phi = \sinh^{3/2} z \; l^{-3/2} \hat{\phi}\: .
\eeq
The action (\ref{act}) is not changed by the conformal transformation,
Eqs.~(\ref{conf}).  The corresponding transformed Lagrangian
looks like
\beq
\label{lag}
{\cal L}=\phi\left( \partial _{z}^2 +\Delta ^{(4)} +\xi_{5}R^{(4)}
\right)\phi\; .
\eeq
where $R^{(4)}=12$.  Since we are interested in the Casimir effect
for the bulk scalar in the AdS background,
we shall use this Lagrangian hereafter.

The one-loop effective potential can be written
as \cite{SSS,NS}
\beq
\label{ef1}
V={1\over 2L {\rm Vol}(M_{4})}\log \det (L_5/ \mu^2) \; ,
\eeq
where $L_5=-\partial _{z}^2 -\Delta ^{(4)} -\xi_{5}R^{(4)}=L_1+L_4 $.
  To calculate the effective potential in Eq.~(\ref{ef1}),
we use $\zeta$-function regularization \cite{eez1, BVW}, as was done in
  Refs.~\cite{GPT1,SSS,NS}. Being precise, the very first step in this
procedure consists in the introduction of a mass parameter in order
to work with dimensionless eigenvalues, thus we should write at
every instance  $L_5/\mu^2$, etc. However, as is often done for the
sake of the simplicity of the notation, we will just keep
in mind the presence of
this $\mu$ factor, to recover it explicitly only in the final formulas.

First, we assume that the eigenvalues of $L_1$ and $L_4$ are of the form
$\lambda^2_{n}, \lambda^2_{\alpha} \; \geq 0\;$  (with $n,\alpha =
1,2,\cdots)$
respectively. In terms of these eigenvalues, $\log \det L_5$
can be rewritten as follows:
\beq
\log \det L_5 = {\rm Tr} \log L_5 ={\rm Tr} \log ( L_1 + L_4)
= \sum_{n,\alpha} \log (\lambda^2_{n} + \lambda^2_{\alpha})
\eeq
Since the $\zeta$-function for an arbitrary operator $A$ is defined by
\beq
\zeta (s | A ) \equiv \sum_{m} (\lambda_{m}^2)^{-s}=\sum _{m}
\e^{-s\log \lambda_{m}^2} \; ,
\eeq
it turns out that ${\rm Tr} \log L_5$
can be rewritten as
\beq
\label{par}
{\rm Tr} \log L_5 = -\partial _{s} \zeta (s|L_5) |_{s=0}\; .
\eeq
Furthermore, the $\zeta$-function is related to the
$\Gamma$-function and heat kernel $K_{t} (A)$:
\beq
\label{gam}
\zeta (s|A) = {1\over \Gamma(s)} \int^{\infty}_{0} dt\; t^{s-1}K_{t}(A),
\quad K_{t}(A)=\sum _{m}\e^{-\lambda_{m}^2 t}\; .
\eeq
$L_1$ is a one-dimensional Laplace operator, and the boundary conditions
result in that the brane separation
$L$ can be taken as the width of a one-dimensional potential well.
As a consequence, the eigenvalues of $L_1$ are given by
\beq
\label{eig}
\lambda^2_{n} = \left( {\pi n \over L} \right)^2 \; ,
\eeq
for finite $L$.

\subsection{The one-brane limit ($L \rightarrow \infty$) \label{Sec1.2}}

The above formula leads to the heat kernel $K_{t}(L_1)$
\beq
\label{l1}
K_{t}(L_1) \sim \sum_{n} e^{-t \left( {\pi n \over L} \right)^2}
\sim \int_{0}^{\infty} dy \e^{-t \left( {\pi y \over L} \right)^2}
= {L \over 2\sqrt{\pi t}},
\eeq
where the large-$L$ limit has been taken, namely, the continuous limit of
$n$.
The heat kernel for $L_5$ is written in terms of  $K_{t}(L_1)$ and
$K_{t}(L_4)$ \cite{BVW}, as
\beq
\label{k5}
K_{t}(L_5)= K_{t}(L_1) K_{t}(L_4)\; .
\eeq
By using Eqs.~(\ref{gam}), (\ref{l1}), and (\ref{k5}), we obtain
$\zeta (s|L_{5})$:
\beq
\zeta (s|L_{5}) &=& {1\over \Gamma(s)} \int^{\infty}_{0} dt t^{s-1}
K_{t}(L_1) K_{t}(L_4) \; ,\nn
&\sim& {L \over 2\sqrt{\pi}}{\Gamma \left( s-{1\over 2} \right)
\over \Gamma(s)} {1\over \Gamma \left( s-{1\over 2} \right)}
\int^{\infty}_{0} dt t^{\left( s-{1\over 2} \right)-1} K_{t}(L_4)
+{\cal O}\left( {1 \over L} \right) \; \nn
&=& {L \over 2\sqrt{\pi}}{\Gamma \left( s-{1\over 2} \right)\over \Gamma(s)}
\zeta \left( s-{1\over 2} | L_4  \right)+{\cal O}\left( {1 \over L}
\right)\; .
\eeq
Combined with Eq.~(\ref{par}),
we obtain the effective potential in the large $L$ limit:
\beq
\label{eff}
V &=& -{1\over 2L {\rm Vol}(M_{4})}\left\{ \zeta '(0|L_5/\mu^2) +
\ln \mu^2 \zeta(0|L_5/\mu^2) \right\} \nn
&=& {1\over 2L {\rm Vol}(M_{4})}\zeta \left(-{1\over 2} | L_4/\mu^2 \right)
+ {\cal O}\left( {\mu^2 \over L} \right)\; .
\eeq
Note that the $\mu^2$ factor has to be taken into account for obtaining the
derivative and, as discussed before, it is in fact everywhere present in
each Lagrangian and its eigenvalues (although it is usually not written down
in order to simplify the notation).
For the spherical brane $S_4$ whose radius is ${\cal R}$,
the four-dimensional zeta function $\zeta \left( s| L_4 \right)$ is given by
\beq
\zeta \left( s | L_4 \right) ={{\cal R}^{2s} \over 3}
\left[ \zeta_{H}\left( 2s-3 , {3\over 2} \right)-{1\over 4}
\zeta_{H}\left( 2s-1 , {3\over 2} \right)  \right]\; ,
\eeq
Here we used a Hurwitz zeta-function and a Bernoulli polynomial
as in Ref.\cite{SSS}.
This equation leads to
\beq
\zeta \left( -{1\over 2}| L_4 \right)
= {1\over 3 {\cal R}}
\left[ \zeta_{H}\left( -4 , {3\over 2} \right)-{1\over 4}
\zeta_{H}\left( -2 , {3\over 2} \right) \right]= 0\; . \label{23}
\eeq
As a result, the effective potential  Eq.~(\ref{eff})
becomes zero (as first has been observed in \cite{SSS} and has been
confirmed in \cite{NS}) as $L \to \infty$.  This situation corresponds to
the case of a one-brane configuration.

\subsection{Small distance expansion \label{Sec1.3}}

Using the power of the zeta regularization formulas \cite{eez1,eez1a},
a much more precise (albeit involved) calculation can be carried out
which respects at every step the complete structure of the five-dimensional
zeta function. That is, the full zeta function is preserved till the end,
and
the final expression is given in terms of an expansion on the brane distance
$L$ over the brane compactification radius $\cal R$,
valid for $L/{\cal R} \leq 1$, which complements the one for large brane
distance obtained above. A detailed calculation follows.

As to the specific zeta formulas employed, adhering to
 the classification that has been given in \cite{eez1a},
the case at hand is indeed to be found there (even if at
first sight it would not seem so). It
 corresponds to a two-dimensional quadratic plus
linear form with truncated spectrum. In fact, this is clear
 from the structure of the spectrum yielding the zeta function
\beq
\zeta (s|L_5) =  \mu^{-2s}\sum_{n,l=0}^\infty
(\lambda_n^2+\lambda_l^2)^{-s},
\eeq
where $\mu$ is a dimensional regularization scale that renders the
argument of the zeta function dimensionless.
In the case of the four-dimensional spherical brane of
radius $\cal R$ considered above, this reduces to
\beq
\zeta (s|L_5) &=& \frac{\mu^{-2s}}{6}   \sum_{n,l=0}^\infty
(l+1)(l+2)(2l+3) \nn
&& \times \left( \left( {\pi n \over L} \right)^2
+{\cal R}^{-2} \left(
 l^2+3l+{9\over 4} \right)\right)^{-s}.
\eeq
This zeta function looks awkward, at first sight. But after some reshuffling
it can be brought to exhibit the standard structure mentioned. Specifically,
\beq
\zeta (s|L_5) &=& \frac{{\cal R}^{2s}}{6\mu^{2s}}
\sum_{n,l=0}^\infty
2  \left( l+{3 \over 2} \right) \left[ \left( \left( l+{3 \over 2} \right)^2
+
{\pi^2n^2 {\cal R}^2\over L^2}\right)^{1-s} \right.  \nn
&& \hspace*{20mm} - \left.
\left( {\pi^2n^2 \over L^2}+
{1 \over 4}\right)\left( \left( l+{3 \over 2} \right)^2 +
{\pi^2n^2 {\cal R}^2\over L^2}\right)^{-s}\right] \nn
&\equiv& \frac{{\cal R}^{2s}}{6\mu^{2s}} \left[ Z_1(s) +Z_2(s)\right],
\eeq
where both $Z_1(s)$ and $Z_2(s)$ are obtained by taking derivatives
(see \cite{eedz1} for a discussion of this issue, nontrivial when asymptotic
expansions are involved), with
respect to $x$ at $x=3/2$, of a zeta function of the class just mentioned,
e.g.
\beq
\sum_{l=0}^\infty \left( \left( l+ x \right)^2 + q
\right)^{-s}, \qquad q \equiv {\pi^2n^2 {\cal R}^2\over L^2}.
\eeq
In Refs.~\cite{eez1a}, explicit formulas for the analytical
continuation of this class of zeta functions are given. To be brief (and
forgetting for the moment about the $n$-sum, for simplicity), we just
have to recall the useful asymptotic expansion
\beq
&&\sum_{n=0}^\infty  \left[ (n+c)^2 +q \right]^{-s} \nn &&
\hspace*{8mm} \sim
\left( {1 \over 2} -c \right) q^{-s} +\frac{q^{-s}}{\Gamma (s)}
\sum_{n=1}^\infty \frac{(-1)^n
\Gamma (n+s)}{n!} q^{-n} \zeta_H (-2n,c) \nn &&
\hspace*{12mm} +\frac{\sqrt{\pi}\,
\Gamma (s-1/2)}{2\Gamma (s)} q^{1/2-s} \\
&& \hspace*{12mm} + \frac{2\pi^s}{\Gamma (s)}q^{1/4-s/2}
\sum_{n=1}^\infty n^{s-1/2} \cos (2\pi n c)
K_{s-1/2} \left( 2\pi n \sqrt{q} \right). \nonumber
\eeq
After some calculations, we get for  $Z_1(s)$ and $Z_2(s)$
\beq
Z_1(s) &=& -\frac{1}{2-s} \left( \frac{\pi^2{\cal R}^2}{L^2}
\right)^{2-s}\zeta (2s-4)-\frac{1}{\Gamma (s-1)}
\sum_{n=1}^\infty  \frac{(-1)^n  \Gamma (n+s-2)}{n!}
\nn  &&  \hspace*{18mm} \times
\left( \frac{\pi^2{\cal R}^2}{L^2} \right)^{2-n-s} \zeta (2s+2n-4)
 \zeta_H' (-2n,3/2), \\
Z_2(s) &=& \frac{1}{1-s} \left( \frac{\pi^2{\cal R}^2}{L^2}
\right)^{1-s} \left[ \frac{\pi^2{\cal R}^2}{L^2}\zeta (2s-4)
+\frac{1}{4} \zeta (2s-2) \right] \nn
&& + \frac{1}{\Gamma (s)}\sum_{n=1}^\infty
\frac{(-1)^n  \Gamma (n+s-1)}{n!}
\left( \frac{\pi^2{\cal R}^2}{L^2} \right)^{1-n-s} \\
&& \hspace*{5mm} \times
\left[ \frac{\pi^2{\cal R}^2}{L^2}\zeta (2s+2n-4)
+\frac{1}{4} \zeta (2s+2n-2) \right] \zeta_H' (-2n,3/2). \nonumber
\eeq

Finally, for the derivative of the five-dimensional zeta function
at $s=0$, we obtain
\beq
\label{twofin}
\zeta' (0|L_5) &=& \frac{\zeta' (-4)}{6}\,
\frac{\pi^4 {\cal R}^4}{L^4} +
 \frac{\zeta' (-2)}{12}\, \frac{\pi^2 {\cal R}^2}{L^2} \nn
&& + \frac{1}{24} \left[  \zeta_H' (-4,3/2)
 - \frac{1}{2} \zeta_H' (-2,3/2)\right]
\ln \frac{\pi^2 {\cal R}^2}{L^2}  \nn  && + \frac{\zeta' (0)}{6}
 \left[  \zeta_H' (-4,3/2) - \frac{1}{2} \zeta_H' (-2,3/2)\right] +
\frac{1}{24} \zeta_H' (-4,3/2)  \nn  && +
\frac{1}{36} \left[ \frac{1}{8}   \zeta_H' (-4,3/2) - \frac{1}{3}
\zeta_H' (-6,3/2)\right] \frac{L^2}{{\cal R}^2} + {\cal O} \left(
\frac{L^4}{
\pi^4 {\cal R}^4}\right)  \nonumber \\ && \hspace*{-6mm} \simeq 0.129652 \,
\frac{{\cal R}^4}{L^4}
- 0.025039 \, \frac{{\cal R}^2}{L^2} - 0.002951 \, \ln
\frac{{\cal R}^2}{L^2} \nn
&&  -0.017956 - 0.000315 \frac{L^2}{{\cal R}^2} +
\cdots \label{131}
\eeq

\subsection{The dynamics of the brane \label{Sec1.4}}

We now consider the dynamics of the dS brane,
which is taken to be the four-dimensional sphere S$_4$, as in
Ref.~\cite{SSS}.  The bulk part is given by  five-dimensional
Euclidean Anti-de Sitter space, Eq.~(\ref{met1}),
which can be rewritten as
\be
\label{AdS5i}
ds^2_\AdS5=dy^2 + l^2\sinh^2 {y \over l}d\Omega^2_4\ .
\ee
One also assumes that the boundary (brane) lies at $y=y_0$
and the bulk space is obtained by gluing two regions,
given by $0 \leq y < y_0$ (see \cite{HHR} for more details.)

We start with the action $S$ which is the sum of
the Einstein-Hilbert action $\SEH$, the Gibbons-Hawking
surface term $\SGH$ \cite{GH}, and the surface counter-term $S_1$, e.g.
\beq
\label{Stotal}
S&=&\SEH + \SGH + 2 S_1  \\
\label{SEHi}
\SEH&=&{1 \over 16\pi G}\int d^5 x \sqrt{\gfv}\left(R_{(5)}
+ {12 \over l^2}\right) \\
\label{GHi}
\SGH&=&{1 \over 8\pi G}\int d^4 x \sqrt{\gfr}\nabla_\mu n^\mu \\
\label{S1}
S_1&=& -{3 \over 8\pi G l}\int d^4 x \sqrt{\gfr} \; .
\eeq
Hereafter the quantities in the  five-dimensional bulk spacetime are
specified by the subindices $_{(5)}$ and those in the boundary
four-dimensional spacetime are by $_{(4)}$.
The factor $2$ in front of $S_1$ in (\ref{Stotal}) is coming from
the fact that we have two bulk regions, which
are connected with each other by the brane.
In (\ref{GHi}), $n^\mu$ is the unit vector normal to the boundary.

If we change the coordinate $\xi$ in
the metric of $\S4$, Eq.~(\ref{ss4}), to $\sigma$ by
\be
\label{S4chng}
\sin\xi = \pm {1 \over \cosh \sigma} \ ,
\ee
we obtain
\be
\label{S4metric2}
d\Omega^2_4= {1 \over \cosh^2 \sigma}\left(d \sigma^2
+ d\Omega^2_3\right)\ .
\ee
For later convenience,
one can rewrite the metric of the five-dimensional space,
Eqs.~(\ref{AdS5i}), (\ref{S4metric2}), as follows:
\be
\label{metric1}
ds^2=dy^2 + \e^{2A(y,\sigma)}\tilde g_{\mu\nu}dx^\mu dx^\nu\ ,
\quad \tilde g_{\mu\nu}dx^\mu dx^\nu\equiv l^2\left(d \sigma^2
+ d\Omega^2_3\right).
\ee
 From Eq.~(\ref{metric1}), the actions (\ref{SEHi}),
(\ref{GHi}), (\ref{S1}), have the following forms:
\beq
\label{SEHii}
\SEH&=& {l^4 V_3 \over 16\pi G}\int dy d\sigma \left\{\left( -8
\partial_y^2 A - 20 (\partial_y A)^2\right)\e^{4A} \right. \nn
&& \left. +\left(-6\partial_\sigma^2 A - 6 (\partial_\sigma A)^2
+ 6 \right){\e^{2A} \over l^2}  + {12 \over l^2} \e^{4A}\right\} \\
\label{GHii}
\SGH&=& {4 l^4 V_3 \over 8\pi G}\int d\sigma \e^{4A}
\partial_y A \\
\label{S1ii}
S_1&=& - {3l^3 V_3 \over 8\pi G}\int d\sigma \e^{4A} \; .
\eeq
Here $V_3=\int d\Omega_3$ is the volume (or area) of
the unit three-dimensional sphere.

As it follows from the discussion in the previous subsections,
there is  a gravitational Casimir contribution coming
from bulk quantum fields. As one sees in the simple example of a
bulk scalar, $S_{\rm Csmr}$ (leading term)  has typically the following form
\be
\label{SCsmr}
S_{\rm Csmr}={cV_3 \over {\cal R}^5}\int dy
d\sigma \e^{-A}.
\ee
Here $c$ is some coefficient, whose value and sign depend
on the type of bulk field (scalar, spinor, vector, graviton, ...)
and on parameters of the bulk theory (mass, scalar-gravitational
coupling constant, etc). In a previous subsection we have found this
coefficient for a conformal scalar. For the following discussion
it is more convenient to consider this coefficient to be some
parameter of the theory. Doing so, the results are quite common and
may be applied to an arbitrary quantum bulk theory. We also assume
that there are no background
bulk fields in the theory (except for the bulk gravitational field).

Adding the quantum bulk contribution
to the action $S$ in (\ref{Stotal}), one can regard
\be
\label{StotalB}
S_{\rm total}=S+S_{\rm Csmr}
\ee
as the total action. In (\ref{SCsmr}), ${\cal R}$ is
the radius of S$_4$.

In the bulk, one obtains the following equation of motion
from $\SEH + S_{\rm Csmr}$ by  variation over $A$:
\beq
\label{eq1}
0&=& \left(-24 \partial_y^2 A - 48 (\partial_y A)^2
+ {48 \over l^2}\right)\e^{4A} \nn
&& + {1 \over l^2}\left(-12 \partial_\sigma^2 A
- 12 (\partial_\sigma A)^2 + 12\right)\e^{2A}
+ {16\pi G c \over {\cal R}^5 l^4 }\e^{-A}\ .
\eeq
%%%%%%
Let us discuss the solution in the situation when the scale factor depends
on both coordinates: $y$ and $\sigma$.
In Ref.~\cite{SSS}, the solution of (\ref{eq1}) given by an expansion
with respect to $\e^{-{y \over l}}$ was found
by assuming that ${y \over l}$ is large:
\be
\label{S4sl}
\e^A={\sinh {y \over l} \over \cosh \sigma}
- {32\pi Gc l^3 \over 15 {\cal R}^5}\cosh^4\sigma
\e^{-{4y \over l}} + {\cal O}\left(\e^{-{5y \over l}}\right)
\ee
for the perturbation from the solution where the brane
is S$_4$.
On the brane at the boundary,
one gets the following equation:
\beq
\label{eq2}
0 = \left( \partial_y A - {1 \over l} \right)\e^{4A} .
\eeq
Substituting the solutions (\ref{S4sl})
into (\ref{eq2}), we find that
\be
\label{S4slbr2}
0 \sim \left( {1 \over {\cal R}}
\sqrt{1 + { {\cal R}^2 \over l^2 }}
 + {2 \pi G l^2 c \over 3 {\cal R}^{10}}\cosh^{5} \sigma
 - {1 \over l} \right) \; .
\ee
Eq.~(\ref{S4slbr2}) tells us that
the Casimir force deforms the shape of S$_4$, since
${\cal R}$ depends on $\sigma$.
The effect becomes larger for large $\sigma$. In the case of a
S$_4$ brane, the effect becomes large if the distance from the
equator becomes large, since $\sigma$ is related to the angle
coordinate $\xi$ by (\ref{S4chng}). In particular, at the north and
south poles ($\xi=0$, $\pi$), $\cosh\sigma$ diverges and then
${\cal R}$ should vanish. This is not coordinate
singularity. In fact, when $\sigma\to \pm \infty$, the 5d scalar
curvature behaves as
\be
\label{crv5}
R_{(5)}\sim - {20 \over l^2} + {12\pi Gc l \over {\cal R}^5}
\e^{7|\sigma|}\e^{-{7y \over l}} + {\cal O}\left(\e^{-{9y \over l}}
\right)\ .
\ee
This  only tells, however, that the perturbation with respect to
$c$ or $\e^{-{y \over l}}$ breaks down. In fact, when $\sigma$ is
large, the corrections appear in the combination of the power of
$\e^{|\sigma|}\e^{-{y \over l}}$. Then the singularity at the poles
is not real one but if we can sum up the correction terms
in all orders with respect to $\e^{-{y \over l}}$, the singularity
would vanish. Then  we have demonstrated that bulk quantum effects
do have the tendency to support the creation of a de Sitter
brane-world Universe.

The original Euclidean 5d AdS space has a isometry of $SO(5,1)$,
which is identical with the Euclidean 4d conformal symmetry.
The existence of the S$_4$ brane breaks the isometry into
$SO(5)$ rotational symmetry, which makes S$_4$ invariant.
If there is no the Casimir effect, Eq.(\ref{S4slbr2}) has
the $SO(5)$ symmetry. The result in (\ref{S4slbr2}) seems to
indicate that the Casimir force breaks the $SO(5)$ symmetry.
We should note that the effective action  (\ref{SCsmr}) does not
seem to be invariant under the rotational symmetry since the action
seems to depend on the choice of the axis connecting the north and
south poles although the calculation of the Casimir effect should be
invariant under the $SO(5)$ symmetry. Since the Casimir effect is
the non-local effect, the exact form of the effective action should
be non-local. Then more exact form of the effective action might be
obtained, for example, by averaging the action (\ref{SCsmr}) with
respect to the choice of the axis. Such a symmetry can be, in
general, broken spontaneously as in (\ref{S4slbr2}). The breakdown
would occur by choosing the time direction to be parallel with
the axis. Then the $SO(5)$ symmetry is broken to $SO(4)$, which
preserves the rotations making the axis, that is, also north and
south poles, invariant.

We now consider the case when the bulk quantum effects are the leading
ones. From
Eq.~(\ref{S4slbr2}), one obtains
\be
\label{S4slbrC}
{\cal R}^8\sim -{4 \pi G l c \over 3}\cosh^5 \sigma \; .
\ee
Here we only consider the leading term with
respect to $c$, which corresponds to the large ${\cal R}$
approximation.
Thus, we have demonstrated that bulk quantum effects do not violate
(in some cases they even support) the  creation of a de Sitter
brane living in a five-dimensional AdS background.

\subsection{Dynamics of two branes  \label{Sec1.4b}
at small distance}

In this subsection, we consider the dynamics of two dS branes
when the distance between them is small. Before including the Casimir
effect, we consider the following actions.
\beq
\label{AA1}
S&=&\SEH + \sum_{a=\pm}\left(\SGH + 2 S_1\right)  \\
\label{AASEHi}
\SEH&=&{1 \over 16\pi G}\int d^5 x \sqrt{\gfv}\left(R_{(5)}
+ {12 \over l^2}\right) \\
\label{AAGHi}
\SGH^\pm &=&\pm {1 \over 8\pi G}\int d^4 x \sqrt{\gfr}\nabla_\mu n^\mu \\
\label{AAS1}
S_1^\pm&=& \mp{3 \over 8\pi G l^\pm}\int d^4 x \sqrt{\gfr} \; .
\eeq
Here the index $a=\pm$ distinguishes the two branes and
we assume that the radius ${\cal R}^+$ (${\cal R}^-$)
corresponds to the larger (smaller) brane. The bulk space is
AdS again and, on the branes, we obtain the following equations:
\be
\label{AA2}
{1 \over {\cal R}^\pm}
\sqrt{1 + { {{\cal R}^{\pm}}^2 \over l^2 }} = {1 \over l^\pm}\ .
\ee
The left-hand side in (\ref{AA2}) is a monotonically
decreasing function with respect to ${\cal R}$.
Since the left-hand side becomes $+\infty$ when ${\cal R}\to 0$
and ${1 \over l}$ when ${\cal R}\to +\infty$, there is a solution
when
\be
\label{AA3}
l>l^+>l^-\ .
\ee

We now include the Casimir effect. First, we consider the backreaction
to the bulk geometry. As we assume the distance between the branes
is small, the radius of the branes are almost constant. The
distance $L$ in (\ref{twofin}) is given by
$\left|z^+ - z^-\right|$, the energy
density by the Casimir effect would be proportional to
${\e^{-5A} \over L^5}$. Then the effective action would be
\be
\label{AASCsmr}
S_{\rm Csmr}={\tilde c V_3 \over L^5}\int dy
d\sigma \e^{-A}.
\ee
Therefore, as in the previous section, the bulk geometry would be
deformed as
\be
\label{AAS4sl}
\e^A={\sinh {y \over l} \over \cosh \sigma}
- {32\pi G\tilde c l^3 \over 15 L^5}\cosh^4\sigma
\e^{-{4y \over l}} + {\cal O}\left(\e^{-{5y \over l}}\right)\ .
\ee
In this case, the equation of the brane corresponding to (\ref{S4slbr2}),
has the following form
\be
\label{AAS4slbr2}
0 \sim \left( {1 \over {\cal R}^\pm}
\sqrt{1 + { {{\cal R}^\pm}^2 \over l^2 }}
 \pm {2 \pi G l^2 \tilde c \over 3 L^{10}}\cosh^{5} \sigma
 - {1 \over l^\pm} \right) \; .
\ee
Eq.~(\ref{AAS4slbr2}) tells us that
the Casimir force deforms the shape of S$_4$ and
the effect becomes larger for large $\sigma$, again, as
in the previous section. We should note, however, the signs
of the contribution from the Casimir effect are different for
the larger and smaller branes. Then if the radius of the larger
brane becomes large (small), that in the smaller one it becomes
small (large). It is interesting that if larger brane is physical universe,
this may serve as dynamical mechanism of decreasing of the cosmological
constant.

\subsection{Stabilization of the radion potential\label{Sec1.5}}

In this subsection, we consider the
stabilization of the radion potential following Ref.~\cite{GLW}.
As first setup, we prepare the suitable metric
and action for the discussion of the stabilization of the radion potential.
\be
\label{met01}
ds^2 = \e^{-2kr_c|\phi|} \eta _{\mu\nu} dx^{\mu}dx^{\nu}
- r_c^2  d\phi^2
\ee
Here $\phi$ is the coordinate on five-dimensions
and $x^{\mu}$ are the coordinates on the four-dimensional
surfaces of constant $\phi$, and $-\pi \leq \phi \leq \pi$
with $(x,\phi)$ and $(x, -\phi)$ identified.
The coordinate $z$ in the metric (\ref{met1})
corresponds to $\e^{kr_c \phi}/k$ in Eq.~(\ref{met01}), and the distance
between two branes $L$ corresponds to $(\e^{\pi k r_c}-\e^{-\pi k r_c})/k$.

We assume that a potential can arise classically
from the presence of a bulk scalar with interaction terms
that are localized at the two 3-branes.
The action of the model with scalar field $\Phi$ is given by
\beq
\label{radS}
S_b = {1\over 2} \int dx^4  \int^{\pi}_{-\pi}
d\phi \sqrt{G} \left( G^{AB} \partial_{A}\Phi  \partial_{B}\Phi
-m^2 \Phi^2 \right) ,
\eeq
where $G_{AB}$ with $A,B = \mu , \; \phi$ as in Eq.~(\ref{met01}).
The interaction terms on the hidden and visible branes
(at $\phi = 0$ and $\phi =\pi$ respectively) are
also given by
\beq
S_h &=& - \int d^{4} x \sqrt{-g_h} \lambda_h (\Phi^2 -v_h^2 )^2 \; , \\
S_v &=& - \int d^{4} x \sqrt{-g_v} \lambda_v (\Phi^2 -v_v^2 )^2 \; ,
\eeq
where $g_h$ and $g_v$ are the determinants of the induced
metric on the hidden and visible branes respectively.

The general solution for $\Phi$ which only depends on
the coordinate $\phi$ is taken from the equation of motion of
the action with respect to $\Phi$, to have the following form:
\beq
\label{kai}
\Phi (\phi) = \e^{2\sigma} [ A\e^{\nu \sigma} + B \e^{-\nu \sigma} ]  \; ,
\eeq
where $\sigma = kr_c |\phi |$ and $\nu = \sqrt{4+m^2/k^2}$.
Substituting this solution (\ref{kai}) into the action
and integrating over $\phi$ yields an effective four-dimensional
potential for $r_c$ which has the form \cite{GLW}
\beq
\label{pott}
V_{\Phi} ( r_c ) &=& k(\nu +2)A^2 (\e^{2\nu k r_c \pi}-1)
+k(\nu -2) B^2 (1-\e^{-2\nu kr_c \pi} ) \nn
&&+\lambda_{v}\e^{-4kr_c \pi}(\Phi (\pi)^2 -v_v^2 )^2
+\lambda_{h} (\Phi (0)^2 -v_h^2 )^2 \;
\eeq
The unknown coefficients $A$ and $B$ are determined by imposing
appropriate boundary conditions on the 3-branes.
Recalling Ref.~\cite{GLW}, the coefficients $A$ and $B$ are given by
\beq
A &=& v_v \e^{-(2+\nu)kr_c \pi} -v_h \e^{-2\nu kr_c \pi}\; ,\\
B &=& v_h (1 + \e^{-2\nu kr_c \pi} ) -v_v \e^{-(2+\nu)kr_c \pi }\; ,
\eeq
for large $kr_c$ limit.
Here we take $\Phi (0)=v_h$ and $\Phi(\pi) = v_v$.

We now consider the case that $kr_c$ is large and
$m/k \ll 1$ for simplicity as in Ref.\cite{GLW},
so that $\nu = 2 + \epsilon$ with $\epsilon \sim m^2/4k^2$ being a small
quantity. Small $m/k$ should generate correct hierarchy \cite{GLW2}.
We also assume $\epsilon kr_c$ is ${\cal O}(1)$ quantity.
Then the potential (\ref{pott}) becomes
\beq
\label{VGW1}
V_{\Phi}(r_c)&=& k\epsilon v_h^2 +4k \e^{-4kr_c \pi}
\left( v_v -v_h  \e^{-\epsilon kr_c \pi} \right)^2
\left( 1+{\epsilon \over 4} \right) \nn
&&- k \epsilon v_h \e^{-(4+\epsilon )kr_c \pi}
(2 v_v -v_h \e^{\epsilon kr_c \pi} ) \; ,
\eeq
and its minimum is given by
\beq
r_{0} = \left({4 \over \pi }\right){k \over m^2}
\ln \left[ {v_h \over v_v} \right]\; .
\eeq
When $kr_c$ is large, $L=(\e^{\pi k r_c}-\e^{-\pi k r_c})/k \sim
\e^{\pi k r_c}/k $ is also large. Then one may assume that the effective
potential includes the term induced by the Casimir effect
as ${\alpha \over L}$ discussed in subsection \ref{Sec1.2},
where $\alpha$ is some constant.\footnote{Note that a Casimir term may
be induced by other bulk fields.}
Thus, we shall add this term to the
potential (\ref{VGW1}) and consider the first order correction to $r_c$
with respect to $\alpha$. Then by assuming $r_c = r_{0}+\delta r_c$, we
find the minimum of the potential is shifted by
\beq
\delta r_c &=&-{\alpha \; \e^{(3+\epsilon)\pi k r_0}
\over 16 k \pi v_h v_v \epsilon } + {1\over 4k\pi} \nn
&\sim & -{\alpha \; \e^{ 3 \pi k r_0} \over 16 k \pi v_h
v_v \epsilon } \; ,
\eeq
where terms of order $\epsilon ^2$ and the higher order terms
with respect to $\e^{ - \pi k r_0}$ are neglected.
The role of Casimir effect is in only to shift slightly the minimum.

In the small $kr_c$ limit, which corresponds to the small $L$ limit as well,
the coefficients $A$ and $B$ in the radion potential (\ref{pott})
are changed as follows
\beq
A &=& {1\over 2kr_c \pi \nu} \left\{ v_v (1+kr_c \pi (\nu -2) ) - v_h
\right\} \\
B &=& {1\over 2kr_c \pi \nu} \left\{ -v_v (1+kr_c \pi (\nu -2) ) +
v_h ( 1+2\nu kr_c \pi ) \right\} \; .
\eeq
In this limit, we suppose that $m/k \gg 1$, so that $\nu \sim {m/k}$,
which makes the situation simple.
The effective potential might include
the term induced by the Casimir effect
as ${\beta \over L^5}$ discussed in subsection \ref{Sec1.3},
where $\beta$ is some constant.
Then, the radion potential in the small $kr_c$ limit is
\beq
\label{ppt}
V_{\Phi} ( r_c ) &=& 2 m r_c \pi k \left( {m\over k} +2 \right) A^2
+2 m r_c \pi k\left( {m\over k} - 2 \right) B^2
+ {\beta \over L^5} \nn
&\sim & { 1 \over  r_c \pi }( v_v - v_h)^2 + {\beta \over (2\pi r_c )^5 } \;
,
\eeq
being here  $L \sim 2\pi r_c$.
To obtain the minimum of the potential,
we differentiate Eq.~(\ref{ppt}) with respect to $r_c$:
\beq
{d \over dr_c }V_{\Phi} ( r_c )
= -{ 1 \over r_c^2 \pi}( v_v - v_h)^2 - {5 \beta \over (2\pi)^5 r_c^6 } \; .
\eeq
Then, if $\beta \leq 0$, the extremum of the potential is reached at
\beq
r_c = \pm {1\over 2 \pi ( v_v - v_h )^{1/2} }
\left( - {5\beta \over 2} \right)^{1/4} \; .
\eeq
%Therefore, the role of the Casimir effect in brane stabilization
%is seen to be {\it essential}.
The extremum is, however, maximum then the stabilization should be local.
Let us give some numbers.
If $v_v$, $v_h\sim
\left(10^{19}{\rm GeV}\right)^{3 \over 2}$ and
$\beta \sim \left(10^{19}{\rm GeV}\right)^{-1}$, we have that
$r_c \sim \left(10^{19}{\rm GeV}\right)^{-1}$ and
$kr_c$ could be of ${\cal O}(1)$. Thus, it is not so unnatural
for the hierarchy problem.

%%%%%%%%%%
For the short $r_c$ case, we may not include the scalar field $\Phi$
in (\ref{radS}) but instead we may include the next-to-leading order of the
effective potential (\ref{twofin}), induced by the Casimir effect,
although the next-to-leading term should be neglected for the
flat brane corresponding to ${\cal R}\to + \infty$:
\be
\label{CaRa}
V_C(r_c)={\beta_1 \over \left(2\pi r_c\right)^5}
+ {\beta_2 \over \left(2\pi r_c\right)^3} \ .
\ee
 From (\ref{twofin}), we see that $\beta_1>0$ and $\beta_2<0$.
As a consequence, in the above potential, there is a minimum at
\be
\label{CARab}
r_c={1 \over 2\pi}\sqrt{-{5\beta_1 \over 3 \beta_2}}\ \simeq
0.4675l\ .
\ee
%%%
The result in (\ref{twofin}) is not for flat brane but
for de Sitter brane and only including the contribution
from massless scalar. We also put a length parameter $l$
in (\ref{CARab}). Then the numerical value in (\ref{CARab})
would be changed but hopefully the main structure would not
be changed.
%%%
We conclude, therefore, that with the only consideration
of the Casimir
effect, the brane might get stabilized, which is a nice
result.\footnote{Note however that thermal effects \cite{BMNO}
may significantly change
the above discussion.}

As we will see later in (\ref{masscorrection}), when one
considers the massive scalar with small mass, there appears
the correction to the effective potential. Motivated with
such result, one considers the following correction to the
effective potential, which corresponds to the leading
term in (\ref{masscorrection}) when $L$ is small:
\be
\label{CRA1}
\Delta V_C(r_c)= {\beta_3 m^2 \over 2\pi r_c}\ .
\ee
Here $m$ expresses the mass of the scalar field.
The result in (\ref{masscorrection}) suggests that $\beta_3$
is negative. By assuming that the correction term (\ref{CRA1}) is
dominant compared with the third (logarithmic) term in
(\ref{twofin}), the minimum in (\ref{CARab}) is shifted as
\be
\label{CARaS}
r_c={1 \over 2\pi}\sqrt{-{5\beta_1 \over 3 \beta_2}}
\left(1 + {5\beta_1 \beta_3 m^2 \over 18 \beta_2^2}
+ {\cal O}\left(m^4\right) \right)\ .
\ee
Then the contribution from small mass has a tendency to
make the distance between the two branes smaller.

%%%%%%%%%%%

\section{Casimir effect for the de Sitter brane \\
in a five-dimensional de Sitter background}

\subsection{Effective potential for the brane}

Next, we use the Euclideanised form of the five-dimensional de Sitter (dS)
metric for a four-dimensional dS brane as follows:
\beq
\label{met}
ds^2 &=& l^2 \left( d\theta^2 +\sin^2 \theta d\Omega _{4}^2 \right), \\
&=& {l^2 \over \cosh^2 z} \left( dz^2 + d\Omega _{4}^2 \right),\nonumber
\eeq
where $l$ is the dS radius, which is related to the cosmological constant of
the dS bulk.

We place two dS branes ---which are four-dimensional spheres, as in the
 AdS bulk case--- in a dS background as the one depicted in Fig.~1.
Since the parameter $\theta$ in Eq.~(\ref{met})
takes  values between $0$ and $\pi$, the parameter
$z$ takes  values between $-\infty$ and $\infty$.
As in the AdS bulk case, the distance between the branes can be
defined as $L=|z_{1}- z_{2}|$.  When  $z_{2}$ is
placed at $\infty$, namely $L=\infty$, the two-brane
configuration becomes a one-brane configuration, as seen in
Fig.~1.

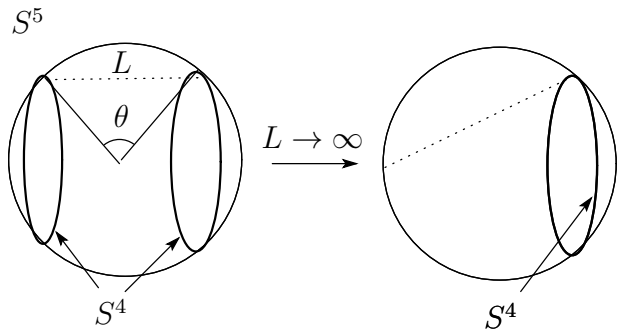
\begin{figure}[htbp]
\begin{center}
%WinTpicVersion2.15
\unitlength 0.1in
\begin{picture}(33.95,15.45)(3.95,-21.30)
% CIRCLE 2 0 3 0
% 4 1220 1800 1210 1190 1210 1190 1210 1190
%
\special{pn 8}%
\special{ar 1220 1400 610 610  0.0000000 6.2831853}%
% ELLIPSE 1 0 3 0
% 4 3560 1830 3690 1360 3570 1360 3570 1360
%
\special{pn 13}%
\special{ar 3560 1430 130 470  0.0000000 6.2831853}%
% ELLIPSE 1 0 3 0
% 4 790 1800 690 1360 790 1370 790 1370
%
\special{pn 13}%
\special{ar 790 1400 100 440  0.0000000 6.2831853}%
% LINE 2 0 3 0
% 2 1200 1800 1600 1330
%
\special{pn 8}%
\special{pa 1200 1400}%
\special{pa 1600 930}%
\special{fp}%
% LINE 2 0 3 0
% 2 1190 1820 800 1370
%
\special{pn 8}%
\special{pa 1190 1420}%
\special{pa 800 970}%
\special{fp}%
% LINE 2 2 3 0
% 2 810 1380 1600 1370
%
\special{pn 8}%
\special{pa 810 980}%
\special{pa 1600 970}%
\special{dt 0.045}%
\special{pa 1600 970}%
\special{pa 1599 970}%
\special{dt 0.045}%
% STR 2 0 3 0
% 3 1200 1160 1200 1260 5 0
% $L$
\put(12.0000,-8.9000){\makebox(0,0){$L$}}%
% STR 2 0 3 0
% 3 710 970 710 1070 5 0
% $S^{5}$
\put(7.1000,-6.7000){\makebox(0,0){$S^{5}$}}%
% CIRCLE 2 0 3 0
% 4 1190 1820 1270 1720 1270 1710 1110 1720
%
\special{pn 8}%
\special{ar 1190 1420 128 128  4.0376480 5.3411853}%
% STR 2 0 3 0
% 3 1200 1470 1200 1570 5 0
% $\theta$
\put(12.0000,-11.7000){\makebox(0,0){$\theta$}}%
% STR 2 0 3 0
% 3 1245 2495 1245 2595 5 0
% $S^{4}$
\put(11.4500,-21.9500){\makebox(0,0){$S^{4}$}}%
% VECTOR 2 0 3 0
% 2 1220 2490 1500 2210
%
\special{pn 8}%
\special{pa 1220 2090}%
\special{pa 1500 1810}%
\special{fp}%
\special{sh 1}%
\special{pa 1500 1810}%
\special{pa 1439 1843}%
\special{pa 1462 1848}%
\special{pa 1467 1871}%
\special{pa 1500 1810}%
\special{fp}%
% VECTOR 2 0 3 0
% 2 1080 2470 860 2180
%
\special{pn 8}%
\special{pa 1080 2070}%
\special{pa 860 1780}%
\special{fp}%
\special{sh 1}%
\special{pa 860 1780}%
\special{pa 884 1845}%
\special{pa 892 1822}%
\special{pa 916 1821}%
\special{pa 860 1780}%
\special{fp}%
% CIRCLE 2 0 3 0
% 4 1220 1800 1210 1190 1210 1190 1210 1190
%
\special{pn 8}%
\special{ar 1220 1400 610 610  0.0000000 6.2831853}%
% ELLIPSE 1 0 3 0
% 4 3560 1830 3690 1360 3570 1360 3570 1360
%
\special{pn 13}%
\special{ar 3560 1430 130 470  0.0000000 6.2831853}%
% STR 2 0 3 0
% 3 2370 1580 2370 1680 5 0
% $L\to \infty$
\put(22.0000,-12.8000){\makebox(0,0){$L\to \infty$}}%
% STR 2 0 3 0
% 3 710 970 710 1070 5 0
% $S^{5}$
%\put(7.1000,-6.7000){\makebox(0,0){$S^{5}$}}%
% STR 2 0 3 0
% 3 1245 2495 1245 2595 5 0
% $S^{4}$
%\put(12.4500,-21.9500){\makebox(0,0){$S^{4}$}}%
% VECTOR 2 0 3 0
% 2 3290 2490 3660 2000
%
\special{pn 8}%
\special{pa 3290 2090}%
\special{pa 3660 1600}%
\special{fp}%
\special{sh 1}%
\special{pa 3660 1600}%
\special{pa 3604 1641}%
\special{pa 3628 1643}%
\special{pa 3636 1665}%
\special{pa 3660 1600}%
\special{fp}%
% LINE 2 2 3 0
% 2 2580 1840 3570 1360
%
\special{pn 8}%
\special{pa 2580 1440}%
\special{pa 3570 960}%
\special{dt 0.045}%
\special{pa 3570 960}%
\special{pa 3569 960}%
\special{dt 0.045}%
% CIRCLE 2 0 3 0
% 4 3180 1820 3170 1210 3170 1210 3170 1210
%
\special{pn 8}%
\special{ar 3180 1420 610 610  0.0000000 6.2831853}%
% CIRCLE 2 0 3 0
% 4 3180 1820 3170 1210 3170 1210 3170 1210
%
\special{pn 8}%
\special{ar 3180 1420 610 610  0.0000000 6.2831853}%
% ELLIPSE 1 0 3 0
% 4 1590 1810 1720 1340 1600 1340 1600 1340
%
\special{pn 13}%
\special{ar 1590 1410 130 470  0.0000000 6.2831853}%
% STR 2 0 3 0
% 3 3185 2515 3185 2615 5 0
% $S^{4}$
\put(31.8500,-22.1500){\makebox(0,0){$S^{4}$}}%
% STR 2 0 3 0
% 3 3185 2515 3185 2615 5 0
% $S^{4}$
\put(31.8500,-22.1500){\makebox(0,0){$S^{4}$}}%
% VECTOR 2 0 3 0
% 2 1990 1820 2430 1820
%
\special{pn 8}%
\special{pa 1990 1420}%
\special{pa 2430 1420}%
\special{fp}%
\special{sh 1}%
\special{pa 2430 1420}%
\special{pa 2363 1400}%
\special{pa 2377 1420}%
\special{pa 2363 1440}%
\special{pa 2430 1420}%
\special{fp}%
\end{picture}%
\end{center}
\caption{The two dS branes are placed in the dS$_5$ background.
The two-brane configuration becomes a
one-brane configuration as $L\to \infty$.}
\end{figure}

The Casimir effect for the bulk scalar in dS background
can be calculated by using the same method as in AdS bulk.

Namely, the Lagrangian for a conformally invariant
massless scalar with scalar-gravitational coupling,
is obtained by  conformal transformation
of the action, Eq.~(\ref{act}), for the metric and the scalar field
 given by
\beq
\label{conf2}
g_{\mu\nu} = \cosh^{-2} z\; l^{2} \hat{g}_{\mu\nu}\; , \quad
\phi = \cosh^{3/2} z \; l^{-3/2} \hat{\phi}\: .
\eeq
Then the Lagrangian is of the same form of Eq.~(\ref{lag}).

The one-loop effective potential is calculated
by means of $\zeta$-function regularization techniques.
Then, the calculated result for
the effective potential in the large $L$ limit
is of the same form of Eq.~(\ref{eff}).
Since the effective potential in Eq.~(\ref{eff})
becomes zero at $L \to \infty$, the effective
potential of the one-brane configuration becomes zero.
Note that this means that the effective potential for $B_5$,
which is the right part in Fig.~1, is zero. Concerning the small
distance expansion, for a potential  corresponding to a conformally
invariant scalar we have an expression as Eq.~(\ref{131}). No
essential difference is encountered in this case.

\subsection{The dynamics of the brane}

The dynamics of  dS brane in a
five-dimensional Euclidean de Sitter bulk
can be considered in a similar way as for the
AdS bulk. The brane is de Sitter, and is taken to be a
four-dimensional sphere S$_4$, as in the previous section.
The five-dimensional Euclidean de Sitter space Eq.~(\ref{met})
can be rewritten as
\be
\label{dS5i}
ds^2_\dS5=dy^2 + \sin ^2 {y \over l} d\Omega^2_4\ .
\ee
Here, we adopt Eq.~(\ref{S4metric2}) for the metric of S$_4$.
We assume that the brane lies at $y=y_0$
and that the bulk is obtained by gluing two regions
given by $0 \leq y < y_0$.

The total action $S$ is the sum of
the Einstein-Hilbert action $\SEH$, the Gibbons-Hawking
surface term $\SGH$, and the surface counter term $S_1$:
like in the AdS bulk case:
\beq
\label{Stotal2}
S = \SEH + \SGH + 2 S_1  \; .
\eeq
The Einstein-Hilbert action $\SEH$ is
\beq
\label{SEHi2}
\SEH = {1 \over 16\pi G}\int d^5 x \sqrt{\gfv}\left(R_{(5)}
- {12 \over l^2}\right)
\eeq
The Gibbons-Hawking
surface term $\SGH$ and the surface counter term $S_1$
are of the same forms as in Eqs.~(\ref{GHi}), (\ref{S1}).

For later convenience, we rewrite the metric of the five-dimensional dS
space, Eqs.~(\ref{dS5i}), (\ref{S4metric2}), as follows:
\be
\label{metric2}
ds^2=dy^2 + \e^{2A(y,\sigma)}\tilde g_{\mu\nu}dx^\mu dx^\nu\ ,
\quad \tilde g_{\mu\nu}dx^\mu dx^\nu\equiv l^2\left(d \sigma^2
+ d\Omega^2_3\right)
\ee
By using Eq.~(\ref{metric2}), the action Eq.~(\ref{SEHi2})
becomes
\beq
\label{SEH2ii}
\SEH&=& {l^4 V_3 \over 16\pi G}\int dy d\sigma \left[\left( -8
\partial_y^2 A - 20 (\partial_y A)^2\right)\e^{4A} \right. \nn
&& \left. +\left(-6\partial_\sigma^2 A - 6 (\partial_\sigma A)^2
+ 6 \right){\e^{2A} \over l^2}  - {12 \over l^2} \e^{4A}\right] \; .
\eeq
which is similar to the AdS bulk case, Eq.~(\ref{SEHii}),
except for the last term. i.e. the cosmological constant.
The Gibbons-Hawking surface term, $\SGH$,
and the surface counter term, $S_1$,
Eqs.~(\ref{GHi}),~(\ref{S1}), have also the same form
of Eqs.~(\ref{GHii}),~(\ref{S1ii}).
We also consider the gravitational Casimir contribution due
to bulk quantum fields. So we add the action of the
Casimir effect, $S_{\rm Csmr}$, ~(\ref{SCsmr})
to the total action $S$  (\ref{Stotal2}).

In the bulk, we obtain the following equation of motion
from $\SEH + S_{\rm Csmr}$ by  variation over $A$:
\beq
\label{eq22}
0&=& \left(-24 \partial_y^2 A - 48 (\partial_y A)^2
- {48 \over l^2}\right)\e^{4A} \nn
&& + {1 \over l^2}\left(-12 \partial_\sigma^2 A
- 12 (\partial_\sigma A)^2 + 12\right)\e^{2A}
+ {16\pi G c \over {\cal R}^5 l^4 }\e^{-A}\ .
\eeq
%%%%%%

For the AdS bulk case, the solution of (\ref{eq22}) can be found
as an expansion with respect to $\e^{-{y \over l}}$,
assuming that ${y \over l}$ is large.
But for the dS bulk case, we cannot adopt the same method,
since the function $\sin {y\over l}$ cannot be
regarded as an expansion with respect to $\e^{-{y \over l}}$.
Thus, we assume the solution to have the following form
\beq
\label{dSass}
\e^A={\sin {y \over l} \over \cosh \sigma} +\delta A \; .
\eeq
Substituting Eq.~(\ref{dSass}) into Eq.~(\ref{eq22}),
we obtain
\beq
\label{eq212}
0 &=& {1\over l^2}\left( -{6\sin{y\over l} \over \cosh \sigma}
+ {\cosh \sigma \over \sin {y\over l} }
- {2 \over \cosh \sigma \sin {y\over l} } \right) \delta A \nn
&& -{4\over l} {\cos {y \over l}\over \cosh \sigma} \partial_{y} (\delta A )
- { 2 \sin {y \over l} \over \cosh \sigma} \partial_{y}^2 (\delta A ) \nn
&& -{\cosh \sigma \over l^2 \sin {y\over l}}
\partial_{\sigma}^2 ( \delta A ) -{4 \pi G c \over  3 {\cal R}^5 l^4}
\left( {\cosh \sigma \over \sin {y\over l}}  \right)^3 \; .
\eeq
We now investigate the behavior of Eq.~(\ref{eq212}) at the north and
south poles ($\xi=0$, $\pi$), that is, as $\cosh\sigma$ diverges.
In this case,  Eq.~(\ref{eq212}) is approximated as
\beq
0 \sim { \e^{\sigma} \over  2 l^2 \sin {y\over l} } \delta A
-{ \e^{\sigma} \over 2 l^2 \sin {y\over l}}
\partial_{\sigma}^2 (\delta A)
-{4 \pi G c \over  3 {\cal R}^5 l^4}
\left( {\e^{\sigma} \over 2 \sin {y\over l}}  \right)^3 \; ,
\eeq
and then
\beq
\label{dSS3}
\delta A - \partial_{\sigma}^2 (\delta A)
\propto { \pi G c \over  3 {\cal R}^5 l^2}
{\e^{2 \sigma} \over  \sin^2 {y\over l} }  \; .
\eeq
Here, we have used the approximation $\cosh\sigma \sim {\e^{\sigma}
\over 2}$. From Eq.~(\ref{dSS3}), we assume
\beq
\label{arbi}
\delta A = \alpha {\e^{2\sigma} \over \sin^2 {y\over l}} \; ,
\eeq
where $\alpha$ is the constant which is obtained by
substituting Eq.~(\ref{arbi}) into Eq.~(\ref{dSS3}), thus
\beq
\alpha = -{\pi G c \over  9 {\cal R}^5 l^2}  \; .
\eeq

The region of the equator $\xi=\pi/2$, namely,
$\cosh\sigma \sim 1 +{1\over 2} \sigma^2 $, Eq.~(\ref{eq212}),
is approximated as
\beq
0&\sim & -\left\{ {1\over l^2}\left( 6\sin{y\over l}
+{2 \over \sin {y\over l} } \right) \delta A \right. \nn
&&\left. +{4\over l}  \cos {y \over l} \partial_{y} (\delta A )
+ 2 \sin {y \over l}  \partial_{y}^2 (\delta A )
\right\} \left( 1 - {1\over 2} \sigma^2 \right) \; .
\eeq

%%%%%%%%%%%%%%%%%%%%%
On the brane at the boundary,
we get the same equation  Eq.~(\ref{eq2}):
\beq
\label{eq21}
0 = \left( \partial_y A - {1 \over l} \right)\e^{4A} .
\eeq
Finally, by substituting the solutions (\ref{dSass})
into (\ref{eq21}), we find
\be
\label{dSee}
0 = {1\over l \cosh \sigma} \left( \cos {y\over l}
-\sin {y\over l} \right) + \partial _{y}(\delta A) \; .
\ee
In the region at the north and south poles, $\cosh \sigma \sim
\e^{|\sigma|}/2$,
if we assume $y={\pi \over 4}l+\delta y$, from Eq.~(\ref{dSee}),
$\delta y$ is obtained by
\beq
\label{dsDlty}
\delta y = {\sqrt{2} \pi G c \over 9 {\cal R}^5 l}
\e^{ 3 |\sigma|} \; .
\eeq
Thus, the deformation of the brane seems to become large at the north and
south pole.

We should note the expression in (\ref{dsDlty}) diverges at
north and south poles where $\sigma\to \pm\infty$.
As in case of AdS bulk in the previous section, this indicates
that the perturbation with respect to $c$ breaks down.
The original Euclidean 5d dS space has a isometry of $SO(6)$,
which is broken by the existence of the S$_4$ brane into $SO(5)$.
Due to the Casimir effect, the $SO(5)$ symmetry seems to be broken to
$SO(4)$, again.

\section{Effective potential for a massive scalar field in the
AdS and dS bulks}

Until now we have dealt with a massless scalar.
In this section we will consider a massive
scalar field in AdS and dS backgrounds.
Let us start with the action for a
massive scalar with scalar-gravitational coupling,
\beq
\label{act3}
{\cal S} = {1\over 2}\int d^{5}x \sqrt{g}
\left[ -g^{\mu\nu}\partial_{\mu} \phi \partial_{\nu} \phi
-m^2 \phi ^2 + \xi_{5}R^{(5)} \phi^2 \right] \; ,
\eeq
For the AdS background with the metric Eq.~(\ref{met1}),
under the conformal transformations (\ref{conf}),
the action changes as
\beq
{\cal S} = {1\over 2}\int d^{5}x \sqrt{g}
\left[ -g^{\mu\nu}\partial_{\mu} \phi \partial_{\nu} \phi
- m^2 l^2 \sinh ^{-2} z \phi ^2 + \xi_{5}R^{(5)} \phi^2 \right] \; ,
\eeq
which yields the Lagrangian for the massive
scalar field with scalar-gravitational coupling
in an AdS background as
\beq
{\cal L}=\phi\left( \partial _{z}^2 +\Delta ^{(4)}
-m^2 l^2 \sinh ^{-2}z + \xi_{5}R^{(4)}
\right) \phi \; .
\eeq
In the above Lagrangian, there appears a singularity at $z=0$.
The point $z=0$ corresponds to $\infty$, where the warp
factor blows up to infinity. Then by putting a brane as the
boundary of the bulk, say putting a brane at $z=z_0<0$ (or $z_0>0$)
and considering the region $z<z_0$ (or $z>z_0$ as bulk space,
the singularity does not appear. And as we can see in
Appendix \ref{A1}, if we include the singular point $z=0$,
a half of the solutions are excluded but there remain other
half of solutions.
%%%%%%%%%%%%%%
 From this Lagrangian, we can calculate the one-loop effective
potential like in the case of a massless scalar field.
The form of the effective potential from the massive scalar
field is given by
\beq
\label{ef2}
V&=&{1\over 2L {\rm Vol}(M_{4})}\log \det (L_5/ \mu^2) \; ,\nn
L_5&\equiv & -\partial _{z}^2 +m^2 l^2 \sinh ^{-2}z -\Delta ^{(4)}
 - \xi_{5}R^{(4)} = L_1 +L_4 \; ,
\eeq
where  the mass term is included in $L_1$.
The eigenvalue of $L_1$ is different from that in Eq.~(\ref{eig}),
 for finite $L$,
since $L_1$ in Eq.~(\ref{ef2}) is
the one-dimensional Schr\"{o}dinger operator
with the potential term $m^2 l^2 \sinh ^{-2}z$.
But this potential term,
which is positive valued and has no bound state,
becomes zero in the limit  $z_2 \to \infty$, that is, when
the distance between branes $L$ becomes $\infty$.
In this case, the eigenvalue of $L_1$ reduces to the
same form of Eq.~(\ref{eig}) and thus the effective potential
becomes zero at the limit of a one-brane configuration.

For the case of a dS background, Eq.~(\ref{met}),
the conformal transformations, Eqs.~(\ref{conf2})
change the action (\ref{act3}) as follows:
\beq
{\cal S} = {1\over 2}\int d^{5}x \sqrt{g}
\left[ -g^{\mu\nu}\partial_{\mu} \phi \partial_{\nu} \phi
- m^2 \cosh ^{-2} z \phi ^2 + \xi_{5}R^{(5)} \phi^2 \right] \; .
\eeq
Then, the Lagrangian for a massive scalar field in the
dS background is given by
\beq
{\cal L}=\phi\left( \partial _{z}^2 +\Delta ^{(4)}
-m^2 \cosh ^{-2}z + \xi_{5}R^{(4)}
\right) \phi \; .
\eeq
Similarly, the effective potential
for the massive scalar field in the dS bulk can be calculated
as in Eqs.~(\ref{ef1}), (\ref{ef2}), by using the operators:
\beq
L_5&\equiv & -\partial _{z}^2 +m^2 \cosh ^{-2}z -\Delta ^{(4)}
 - \xi_{5}R^{(4)} = L_1 +L_4 \; ,
\eeq
where the mass term is included in $L_1$.
The potential term of $L_1$,
$m^2 \cosh ^{-2}z$, has always a positive
value and no bound state like in the AdS case.
It becomes zero in the limit $z_2 \to \infty$ as well.
Therefore, the effective potential for
the massive scalar field in a dS background
also becomes zero in the limit of a one-brane configuration.

\subsection{Small mass limit (with $L$ not large)}

Continuing with the massive scalar field, and for a de Sitter brane in an
AdS
bulk, in the case of the two brane configuration we just need to supplement
the
calculation carried out in Appendix \ref{A1}, which can be done exactly,
with the boundary conditions imposed on the two branes. We thus
obtain a modification
of a perfectly solvable model which appears in several textbooks
(namely, an hyperbolic variant of the celebrated P\"oschl-Teller potential),
albeit with reverse sign and supplemented with the infinite well created
by the branes (as in the massless case).
Since we shall deal with the low and high mass approximations, the
WKB method turns out to be well suited to carry out the analysis.

Setting the branes at
$z=\pm L/2$ (for the sake of symmetry) we get the following results.
In the small mass limit, we obtain a
modification of the eigenvalues of the $L_1$ Lagrangian, in the form
\beq
\lambda_n^2 \simeq {\pi^2n^2 \over \mu^2 L^2} + m^2 l^2
 {\tanh (\mu L/2) \over \mu L/2}.
\eeq
Carrying this into the zeta function, after a further approximation one
gets that the elementary zeta functions in the formulas are modified in
the way, e.g.
\beq
&& \zeta (2s) \rightarrow \zeta (2s) -s \zeta (2s+2) \rho + \frac{s(1+s)}{2}
\zeta (2s+4) \rho^2 + {\mathcal O} (m^6), \nonumber \\
 && \hspace*{4cm}  \rho \equiv
{m^2 l^2 \mu^2L^2 \over \pi^2}\, {\tanh (\mu L/2) \over \mu L/2}.
\eeq
Thus, in the case here considered, when $m$ is small and $L$ is
not very large,
for the derivative of the zeta function at $z=0$ we obtain the following
additional terms ($ l^2 \mu^2=1$):
\beq
\label{masscorrection}
\Delta \zeta' (0|L_5)&\simeq& \frac{a\rho +a^2\rho^2}{48} -\frac{\pi^2}{144}
\left[\frac{a \rho^2}{2} + \left[ 2\zeta'(-4,3/2)-\zeta'(-2,3/2)\right]
\rho\right]\nonumber \\ && -\frac{\pi^4}{4370} \left[ 2\zeta'(-4,3/2)-
\zeta'(-2,3/2)\right]
\rho^2 +{\mathcal O} (m^6), \label{310} \\ && \hspace*{25mm} a\equiv
{\pi^2{\mathcal R}^2 \over L^2}, \quad
\rho \equiv {m^2l^2 \over \pi^2}\, {\tanh ( L/2l) \over
L/2l}.\nonumber
\eeq
These terms have just to be added to the derivative of the zeta function
at $z=0$, Eq.~(\ref{131}), corresponding to the de Sitter brane in AdS
bulk, in order
to obtain
the corresponding effective potential. In a full-fledged analysis of the
different contributions to the effective potential, one has to take into
account the relative importance of the different dimensionless ratios
involved here. The working hypothesis has been that $m^2$ was `small.'
In fact, we see from the final result that $m^2$  most naturally goes with
$l^2$, which also serves as a unit for $L$ and, indirectly, for $\mathcal
R$.
The ordering in Eq.~(\ref{310}) assumes that $a \rho \sim 1$, $\rho < 1$,
but
a lot more information can be extracted from this small-mass expansion.

The calculation in the same case of a massive scalar field but for a de
Sitter brane in a dS bulk (two and one brane configurations) proceeds in a
quite similar fashion. Only, an additional coordinate change is required
at the beguining, to deal with the problem of the singularity of the
potential
of the Schr\"odinger equation at $z=0$ in the initial coordinates, as
carefully explained in the Appendix.

\subsection{Large mass limit (with $L$ not small)}
In this case the calculation turns out
to be more involved. The eigenvalues get modified as follows:
\beq
\lambda_n^2 \simeq {\pi^2n^2 l^2 \over  L^2} +{2
\arctan (\sinh  L/2l) \over \sinh ( L/2l)} \, m^2l^2 +
{ \pi n m l^2 \over L\sinh ( L/2l)} + \cdots
\eeq
However, we will be interested in the dominant contribution only. Thus, in
the approximation which is opposite to the previous one, namely
when $m^2$ is large and $L$ is not very small, we get a simple modification
of
the relevant zeta function, of the form
\beq
\zeta (s|L_5) &=& \frac{L}{2l \sqrt{\pi}}
\frac{\Gamma (s-1/2)}{\Gamma (s)} \nn
&& \times \zeta \left( s-{1 \over 2}\left| L_4  + 2m^2 {\arctan (\sinh (
L/2l))
\over \sinh ( L/2l)}\right.\right) + \cdots
\eeq
And this leads to the following result, for the derivative of the zeta
function at
$z=0$, which is valid for sufficiently large scalar mass and $L$:
\beq
  \zeta' (0|L_5) =- \frac{4 m^2l^3}{3\mathcal R} \,
{\arctan (\sinh  L/2l) \over \sinh  ( L/2l)} + \cdots
\eeq
Again, this is the additional contribution to the derivative of the
full zeta function at $z=0$, the same Eq.~(\ref{23}) but corresponding to
the
de Sitter case.
However, as this derivative was equal to zero in the massless case,
the above expression yields now the {\it whole} value of the derivative and,
 correspondingly, of the effective potential.
Note in fact that this reduces to zero, exponentially fast,
in the one-brane limit ($L \to \infty$), in perfect accordance with
Eq.~(\ref{23}).
Also in this case we are allowed to play with the relative values of the
different dimensionless fractions appearing in our expression.

\subsection{Braneworld stabilization by the Casimir force}

In \cite{NOS}, the brane stabilization via study of radion potential in the
Lorentzian deSitter bulk space was discussed in direct analogy
with AdS case. The branes are spacelike and
the distance between  two branes is time-like and
we denote the distance by $T$.
As in (\ref{CaRa}-\ref{CARaS}), we now consider the contribution
from the Casimir effect to the stabilization. For simplicity,
we do not include the massive scalar field $\Phi$ as in
(\ref{radS}) but we  take the next-to-leading order of the
effective potential (\ref{twofin}), induced by the Casimir effect,
and we assume:
\be
\label{CaRadS}
V_C(T)={\beta^{\rm dS}_1 \over T^5} + {\beta^{\rm dS}_2 \over T^3} \ .
\ee
If $\beta^{\rm dS}_1>0$ and $\beta^{\rm dS}_2<0$ as in (\ref{twofin}),
there is a minimum at
\be
\label{CARabdS}
T=\sqrt{-{5\beta^{\rm dS}_1 \over 3 \beta^{\rm dS}_2}}\ .
\ee
Then even for the branes in the deSitter bulk, only by
the Casimir effect, the brane might get stabilized.

As in (\ref{masscorrection}), when we
consider the Casimir effect from the massive scalar with
small mass, we may consider the following correction to the
effective potential:
\be
\label{CRA1dS}
\Delta V_C(T)= {\beta_3 m^2 \over T}\ .
\ee
Here $m$ expresses the mass of the scalar field.
Then the minimum in (\ref{CARabdS}) is shifted as
\be
\label{CARaSb}
r_c=\sqrt{-{5\beta^{\rm dS}_1 \over 3 \beta^{\rm dS}_2}}
\left(1 + {5\beta^{\rm dS}_1 \beta^{\rm dS}_3 m^2 \over 18
{\beta^{\rm dS}_2}^2}
+ {\cal O}\left(m^4\right) \right)\ .
\ee
Then again the contribution from small mass has a tendency to
make the distance between the two branes smaller.
Thus, the possibility of dS braneworld stabilization occurs in
the same way as with AdS bulk.

%%%%%%%%%%%

\section{Effective potential for a massive scalar
without scalar-gravitational coupling}

In this section we will consider a  more simple case, which
does not include a scalar-gravitational coupling term, $\xi_{5}R^{(5)}
\phi^2$.
The action is simply
\beq
\label{act3b}
{\cal S} = {1\over 2}\int d^{5}x \sqrt{g}
\left[ -g^{\mu\nu}\partial_{\mu} \phi \partial_{\nu} \phi -m^2 \phi ^2
\right] \; .
\eeq
This action is not conformally invariant
under the conformal transformations (\ref{tr1}),
which  change it as
\beq
\label{act4}
{\cal S} &=& {1\over 2}\int d^{5}x \sqrt{\hat g}
\left[ -\e^{3\sigma}\hat g^{\mu\nu}\partial_{\mu}
\left(\e^{-{3 \over 2}\sigma} \hat \phi\right)
\partial_{\nu} \left(\e^{-{3 \over 2}\sigma} \hat\phi\right)
 -m^2 \e^{2\sigma}\hat \phi ^2 \right]\nn
&=& {1\over 2}\int d^{5}x \sqrt{\hat g} \nn
&& \times \left[ -\hat g^{\mu\nu}\partial_{\mu} \hat \phi
\partial_{\nu} \hat \phi - {9 \over 4}
\hat g^{\mu\nu}\partial_{\mu}\sigma
\partial_{\nu} \sigma \hat\phi^2  + 3 \hat \phi\hat g^{\mu\nu}
\partial_{\mu}\sigma \partial_{\nu} \hat\phi
 -m^2 \e^{2\sigma}\hat \phi ^2 \right]\ .
\eeq
where we take $\alpha=2$ and  $\beta =-{3\over 2}$ for simplicity.
The third term in Eq.~(\ref{act4}) can be rewritten as
\be
\label{II}
\hat \phi\hat g^{\mu\nu}
\partial_{\mu}\sigma \partial_{\nu} \hat\phi
={1 \over 2}D^\mu \left(\hat\phi^2\partial_\mu \sigma\right)
 - {1 \over 2}\hat\phi^2\Delta ^{(5)} \sigma
\ee
and using  partial integration, we obtain
\beq
\label{act5}
{\cal S} &=& {1\over 2}\int d^{5}x \sqrt{\hat g}
\left[ -\hat g^{\mu\nu}\partial_{\mu} \hat \phi
\partial_{\nu} \hat \phi - \left({9 \over 4}
\hat g^{\mu\nu}\partial_{\mu}\sigma \partial_{\nu} \sigma
+ {3 \over 2}\Delta ^{(5)}  \sigma\right) \hat \phi^2
 -m^2 \e^{2\sigma}\hat \phi ^2 \right]\ .\nn
\eeq
If we now introduce the AdS background, which has the metric
Eq.~(\ref{met1}),
under the conformal transformations (\ref{conf}),
namely $\e^{2\sigma}=l^2 \sinh^{-2} z$, the action changes as
\beq
{\cal S} &=& {1\over 2}\int d^{5}x \sqrt{g}
\left[ -g^{\mu\nu}\partial_{\mu} \phi \partial_{\nu} \phi
-\left( {9 \over 4} + {15 \over 4}\sinh^{-2} z \right) \phi^2
- m^2 l^2 \sinh ^{-2} z \phi ^2  \right] \; .\nn
\eeq
This action leads the Lagrangian for the massive
scalar field without scalar-gravitational coupling in an AdS background as
\beq
\label{lag5}
{\cal L}=\phi\left( \partial _{z}^2 +\Delta ^{(4)}
-\left( {9 \over 4} + {15 \over 4 }\sinh^{-2} z \right)
-m^2 l^2 \sinh ^{-2}z  \right) \phi \; .
\eeq
Note that the third term in Eq.~(\ref{lag5}),
\beq
\label{zv}
-\left( {9 \over 4} + {15 \over 4 }\sinh^{-2} z \right)\; ,
\eeq
corresponds to
\beq
\label{xx5}
\xi_5\left(R^{(4)} - R^{(5)}\e^{2\sigma}\right) \; ,
\eeq
coming from Eqs.~(\ref{act}), (\ref{lag}),
where $e^{2\sigma}=l^2 \sinh^{-2} z$, because if we put
$\xi_{5}=-3/16$, $R^{(4)}=12,\; R^{(5)}= - {20 \over l^2}$,
which are the scalar curvatures of $S_4$ and AdS$_5$, respectively,
into Eq.~(\ref{xx5}), then Eq.~(\ref{xx5}) coincides with
Eq.~(\ref{zv}) exactly.

The one-loop effective potential can be written as
\beq
\label{effb}
V &=&{1\over 2L {\rm Vol}(M_{4})}\log \det (L_5/ \mu^2) \; ,\nn
L_5 &=& -\partial _{z}^2 -\Delta ^{(4)}
+\left( {9 \over 4} + {15 \over 4 }\sinh^{-2} z \right)
+m^2 l^2 \sinh ^{-2}z  = L_1+L_4 \; ,\nn
L_1 &=& -\partial _{z}^2 + {15 \over 4 }\sinh^{-2} z  +m^2 l^2 \sinh ^{-2}z
,
\quad L_4= {9 \over 4}  -\Delta ^{(4)} \; .
\eeq
Then, the eigenvalue of $L_1$ agrees with Eq.~(\ref{eig})
in the limit when the distance between the two brane becomes
infinity, $L \to \infty$, because the potential terms of (\ref{effb}),
${15 \over 4 }\sinh^{-2} z  +m^2 l^2 \sinh ^{-2}z$, become
zero in this limit.  Therefore, the effective potential for
the massive scalar field without scalar-gravitational coupling
in an AdS background becomes zero in the limit of the one-brane
configuration.

Similarly, the Lagrangian for the massive
scalar field without scalar-gravitational coupling in a dS background
can be seen to be
\beq
{\cal L}=\phi\left( \partial _{z}^2 +\Delta ^{(4)}
-\left( {9 \over 4} - {3 \over 4 }\cosh^{-2} z \right)
-m^2 l^2 \cosh ^{-2}z  \right) \phi \; .
\eeq
In the limit  $L \to \infty$,
the eigenvalue of $L_1$ and the heat kernel $K_{t}(L_1)$
have the same form of Eqs.~(\ref{eig}), (\ref{l1}) as in the AdS case.
Thus, the effective potential  becomes zero too,
in the limit when  the distance between the two branes becomes infinite.

\section{Discussion and conclusions}

To summarize, in this paper we have shown how one can bring the
calculation of the effective potential for a massive or conformal bulk
scalar, in an AdS or dS braneworld  with a dS brane, down to well-known
cases corresponding to zeta-function expansions
 \cite{eez1}. In this way, a complete and detailed analysis of the
different situations can be given, and corrections to the limiting
cases are obtainable at any order.
As our four-dimensional universe is (or will be) in a dS phase, our results
have, potentially, very interesting applications to primordial cosmology.
What is also important, our method and results here open the door to
corresponding  calculations for other quantum fields as spinors, vectors,
 graviton, gravitino, etc.
As we see it, this will only need some more involved calculations, but no
new conceptual problems are expected, at least at the level of the
one-loop efective
potential. In the case of several spin fields, the bulk Casimir
effect may also be found  in this way, at least in principle,  for
supersymmetric theories, including supergravity too.
It is quite possible then, that a five-dimensional AdS gauged
supergravity can be constructed, with AdS being the vacuum state but
still having a dynamically realized de Sitter
brane, which represents our observable universe.

Another issue where bulk quantum effects may play a dominant role
involves moving, curved branes. The corresponding bulk effective
potential might sometimes be a measure of supersymmetry breaking,
and thus be of primordial cosmological importance in the study of the
 very early brane universe.

Finally, the bulk effective potential in realistic SUSY theories
gives a non-trivial contribution to the effective cosmological constant,
in five as well as in four dimensions. Hence, it is conceivable
to use it in a  relaxation of the cosmological constant problem.

\section*{Acknowledgements}

EE is indebted with the Mathematics Department, MIT, and specially with
Dan Freedman for warm hospitality. Very interesting discussions with
Bob Jaffe and collaborators at CTF, MIT, on the Casimir effect
are acknowledged. SDO thanks A. Starobinsky and S. Zerbini for helpful
discussions
on related questions and the IEEC, where this work was initiated, for warm
hospitality. The research by EE is supported in part by DGI/SGPI (Spain),
project BFM2000-0810, and by CIRIT (Generalitat de Catalunya),
contract 1999SGR-00257.
The research by SN is supported in part by the Ministry of
Education, Science, Sports and Culture of Japan under
the grant number 13135208.
The  research by SO is supported in part by the Japanese Society
for the Promotion of Science under the Postdoctoral Research Programme.

\appendix

\section{Appendix
\label{A1}}

\def\AdS5{{{\rm AdS}_5}}
\def\S5{{{\rm S}_5}}

We consider the following Schr\"odinger equation
\be
\label{Ai}
\left(-{d^2 \over dz^2} + {m^2l^2 \over \sinh^2 z} \right)\phi
=\lambda\phi \ .
\ee
This equation is the $z$-dependent part of the Klein-Gordon
equation in $\AdS5$ and $\hat\phi=\sinh^{-{3 \over 2}} z\phi$
corresponds to the original scalar field in the action.
The limit $z=\infty$ corresponds to the infinity in $\AdS5$
at which the warp factor vanishes, and $z=0$ corresponds to the
infinity where the warp factor grows up to infinity.
In (\ref{Ai}) there appears a singularity at $z=0$.
As the point $z=0$ corresponding to $\infty$,
by putting a brane as the boundary of the bulk,
say putting a brane at $z=z_0<0$ (or $z_0>0$),
and considering the region $z<z_0$ (or $z>z_0$) as bulk space,
the singularity does not appear.
%And as we can see in
%Appendix \cite{A1}, the singularity is not real singularity
%for small $m$.

With the redefinitions
\be
\label{Aii}
\phi = \sinh^{1 \over 2} z \psi\ ,\quad x=\cosh z\ ,
\ee
Eq.~(\ref{Ai}) can be rewritten as
\be
\label{Aiii}
0=\left(x^2 -1\right) {d^2 \psi \over dx^2}
+ 2x {d\psi \over dx} - \left(-\lambda - {1 \over 4}
+ {m^2 l^2 + {1 \over 4} \over x^2 -1}\right)\psi \ ,
\ee
whose solutions are given by the associated Legendre functions
$P^{\pm \mu}_\nu (x)$, which are defined in terms of the Gauss
hypergeometric function:
\be
\label{Aiv}
P^\mu_\nu(z)={1 \over \Gamma(1-\mu)} \left({x+1 \over x-1}
\right)^{\mu \over 2}F\left(-\nu, \nu+1, 1-\mu; {1-x \over 2}\right)\ .
\ee
The parameters $\mu$ and $\nu$ are here given by
\be
\label{Av}
\mu^2=l^2m^2 + {1 \over 4}\ ,\quad \nu(\nu+1)=-\lambda - {1 \over 4}\
\mbox{or}\
\nu={-1 \pm \sqrt{ - 4\lambda} \over 2}.
\ee
When $x$ is large, $P^\mu_\nu(x)$ behaves as
\be
\label{Ava}
P^\mu_\nu(x)\sim {1 \over \sqrt{\pi}}\left[
{\Gamma\left(\nu + {1 \over 2}\right) \left(2x\right)^\nu
\over \Gamma\left(\nu - \mu + 1\right) }
+ {\Gamma\left(-\nu+ { 1\over 2}\right) \over
\Gamma\left(-\nu - \mu \right)\left(2x\right)^{\nu + 1}}
\right]\ .
\ee
Since $\phi\sim x^{{1 \over 2}}\psi $,
then in order that $\phi$ is regular there, we have the
constraint that
\be
\label{Avi}
 -4\lambda \leq 0 \ \ \ \mbox{or}\ \ \ \lambda\geq 0\ ,
\ee
which is identical with what we have in the massless case.
When we include the point $z=0$, which corresponds to $x=1$,
when $\sqrt{x-1}\sim z \to 0$, Eq.~(\ref{Aiv}) becomes singular
for positive $\mu$ as $(x-1)^{-{\mu \over 2}}\sim z^{-\mu}$.
As $\phi\sim z^{1 \over 2}\psi \sim z^{{1 \over 2} - \mu}
=z^{{1 \over 2}\left(1 - \sqrt{1 + 4l^2 m^2}\right)}$,
the positive branch of $\mu$ should be excluded and
we must have $\mu=-\sqrt{l^2m^2 + {1 \over 4}}$.

If we do not include the brane, the spectrum for the massive
case is not changed. In order to investigate the effect of
the mass, we put a brane at $x=x_0\gg 1$ (or $z=z_0$).
On the brane, we impose the Neumann boundary condition for $\phi$:
\be
\label{Abr1}
{d\phi \over dz}=0\ ,\quad \left(\Leftrightarrow\
{d\phi \over dx}=0\right)\ .
\ee
For simplicity,
we consider the model where the bulk space includes the
point $x=1$ ($z=0$); hence $\mu=-\sqrt{l^2m^2 + {1 \over 4}}$.
We write $\mu$ and $\nu$ in (\ref{Av}) as
\be
\label{Abr2}
\mu=-\omega - {1 \over 2}\ ,\quad \nu = - {1 \over 2}
+ i\omega \ .
\ee
Then we have $\lambda=\omega^2$. By using (\ref{Ava}), we find,
for large $x$,
\be
\label{Abr3}
\phi(x)\sim {\Gamma (i\omega) \over \Gamma(i\omega + k)}
\left(2x\right)^{i\omega} + {\Gamma (-i\omega) \over
\Gamma(-i\omega + k)} \left(2x\right)^{-i\omega}\ .
\ee
Then the boundary condition (\ref{Abr1}) yields
\be
\label{Abr4}
{\Gamma (i\omega) \over \Gamma(i\omega + k)}
\left(2x_0\right)^{i\omega} - {\Gamma (-i\omega) \over
\Gamma(-i\omega + k)} \left(2x_0\right)^{-i\omega}\ .
\ee
If we assume $\omega$ and $k$ to be small, the Gamma function
can be approximated by $\Gamma (\pm i\omega) \sim \pm
{1 \over i\omega}$ and $\Gamma(\pm i\omega + k) \sim
{1 \over \pm i\omega + k}$. Then, Eq.~(\ref{Abr4}) can be
rewritten as
\be
\label{Abr5}
\ln\left({1 + i{k \over \omega} \over 1 - i{k \over \omega}}
\right) = i\omega \ln \left(2x_0\right) + 2\pi i n\quad
\left(n=0,\pm 1, \pm 2, \cdots\right)\ .
\ee
For large $x_0$, the solution for $n=0$ is given by
\be
\label{Abr6}
\omega \sim {\pi \over \ln \left(2x_0\right)}\ ,
\ee
for non-vanishing $k$ ($m\neq 0$), which gives the following
lower bound for $\lambda$:
\be
\label{Abr7}
\lambda=\omega^2 \geq \left({\pi \over \ln \left(2x_0\right)}
\right)^2 \sim {\pi ^2 \over z_0^2}\ .
\ee

We now consider the equation  for the dS case:
\be
\label{Aib}
\left(-{d^2 \over dz^2} + {m^2l^2 \over \cosh^2 z} \right)\phi
=\lambda\phi \ .
\ee
This equation is the $z$-dependent part of the Klein-Gordon
equation in $\S5$ or Euclidean de Sitter space, and
$\hat\phi=\cosh^{-{3 \over 2}} z\phi$
corresponds to the original scalar field in the action.
The limit of $z=\pm\infty$ corresponds to the south and north
poles in $\S5$.
With the following redefinitions,
\be
\label{Aiib}
\phi = \cosh^{1 \over 2} z \psi\ ,\quad x=\cosh z\ ,
\ee
Eq.~(\ref{Aib}) can be rewritten as
\be
\label{Aiiib}
0=\left(x^2 + 1\right) {d^2 \psi \over dx^2}
+ 2x {d\psi \over dx} - \left(-\lambda - {1 \over 4}
+ {m^2 l^2 + {1 \over 4} \over x^2 + 1}\right)\psi \ .
\ee
If we replace $x$ by $x=iy$, the above equation (\ref{Aiiib}) turns into
\be
\label{Avii}
0=\left(y^2 - 1\right) {d^2 \psi \over dy^2}
+ 2x {d\psi \over dx} - \left(-\lambda - {1 \over 4}
 - {m^2 l^2 + {1 \over 4} \over y^2 - 1}\right)\psi \ .
\ee
Finally, if we choose, as in (\ref{Av}),
\be
\label{Avb}
\mu^2=-\left(l^2m^2 + {1 \over 4}\right)\ ,
\quad \nu(\nu+1)=-\lambda - {1 \over 4}\
\mbox{or}\
\nu={-1 \pm \sqrt{ - 4\lambda} \over 2}\ ,
\ee
the solution of Eq.~(\ref{Avii}) or (\ref{Aiiib}) is given
by the associated Legendre functions $P^{\pm \mu}_\nu (ix)$, again.
Note that $\mu$ in (\ref{Avb}) is imaginary, in general.
Anyhow, in order that $\hat\phi$ be
regular there, we must impose again the same constraint (\ref{Avi}).

\end{document}